\documentclass[aps,prb,twocolumn,superscriptaddress,showpacs]{revtex4}

\usepackage{latexsym,amssymb}
\usepackage{graphicx,color,pstricks}
\usepackage{epsfig,epsf,bm}
\usepackage{color}
\definecolor{gray}{rgb}{0.7,0.7,0.7}

\begin{document}

\title{Nonequilibrium noise correlations in a point contact of helical edge states}
\author{Yu-Wen Lee}
\email{ywlee@thu.edu.tw} \affiliation{Department of Physics,
Tunghai University, Taichung, Taiwan, R.O.C.}
\author{Yu-Li Lee}
\email{yllee@cc.ncue.edu.tw} \affiliation{Department of Physics,
National Changhua University of Education, Changhua, Taiwan, R.O.C.}
\author{Chung-Hou Chung}
\email{chung@mail.nctu.edu.tw} \affiliation{Department of Electrophysics,
National Chiao Tung University, Hsin-Chu, Taiwan, R.O.C.}
\affiliation{Physics Division, National Center for Theoretical Sciences, HsinChu, Taiwan R.O.C. 300}

\begin{abstract}
 We investigate theoretically the nonequilibrium finite-frequency current noise in a four-terminal
 quantum point contact of interacting helical edge states at a finite bias voltage. Special focus is
 put on the effects of the single-particle and two-particle scattering between the two helical edge
 states on the fractional charge quasiparticle excitations shown in the nonequilibrium current
 noise spectra. Via the Keldysh perturbative approach, we find that the effects of the single-particle
 and the two-particle scattering processes on the current noise depend sensitively on the Luttinger
 liquid parameter. Moreover, the Fano factors for the auto- and cross correlations of the currents in
 the terminals are distinct from the ones for tunneling between the chiral edge states in the quantum
 Hall liquid. The current noise spectra in the single-particle-scattering-dominated and the
 two-particle-scattering-dominated regime are shown. Experimental implications of our results on the
 transport through the helical edges in two-dimensional topological insulators are discussed.
\end{abstract}


\maketitle

\section{Introduction}

Ever since the discovery of the quantum Hall effect, there has been a growing interest in topological
properties in certain quantum condensed-matter systems, especially when a model of the topological
states in the absence of applied magnetic fields was constructed.\cite{Hal} Recently, a new topological
state of matter in two dimensions, the quantum spin Hall insulator (QSHI), was theoretically proposed in
various systems with time-reversal symmetry and spin-orbit interactions.\cite{KM,BZ} The hallmark of the
topological nature in QSHIs is the presence of a bulk gap together with gapless edge states.\cite{KM2}
These edge states propagate in opposite directions for opposite spins, and thus are usually dubbed as the
helical liquids.\cite{WBZ} The stability of the helical liquid against the elastic backscattering is
protected by the time-reversal invariance\cite{KM}; hence, the helical liquid forms a distinctive feature
of this new topological state of matter. This state occurs in HgTe/CdTe quantum well structures,\cite{BHZ}
and there has already been experimental evidence in transport properties of helical liquids, which may
be considered as the confirmation of these unique one-dimensional ($1$D) system.\cite{Konig,Brune}

In the presence of electron-electron interactions, these helical edge states form a special type of
Luttinger liquid (LL), the helical LL, in which the spins are associated with the
directions of the momenta.\cite{WBZ} Therefore, it is interesting both theoretically and experimentally
to look for the unique signatures of helical LLs and, in particular, to distinguish them from the the
usual LLs. Recently, it was proposed in Refs. \onlinecite{HKC,TK} that a four-terminal quantum point
contact (QPC) in the QSHI can be used as a probe of the helical LL. In particular, in Ref. \onlinecite{HKC},
it was noted that the problem of the QPC in a QSHI can be mapped onto the model of a
spinful LL with a weak tunneling link. The corresponding LL parameter of the charge mode $K_c=K$ is the
inverse of that of the spin mode $K_s=1/K_c=1/K$. Therefore, the edge states of the QSHI with a tunnel
junction can realize phases which cannot exist for the spin-SU(2) invariant LL with a single impurity
where $K_s=1$ there.\cite{KFFN} It was further shown in Ref. \onlinecite{HKC} that there exists a quantum
critical point which can be tuned by adjusting the value of the gate voltage.\cite{TK} As a result, the
low temperature zero-bias conductance can be described by a universal scaling function of the temperature
and the gate voltage. Later, a duality relation between the charge and spin sectors in such a four-terminal
setup was found in the nonequilibrium situation.\cite{LBRT}

\begin{figure}
\begin{center}
 \includegraphics[width=0.65\columnwidth]{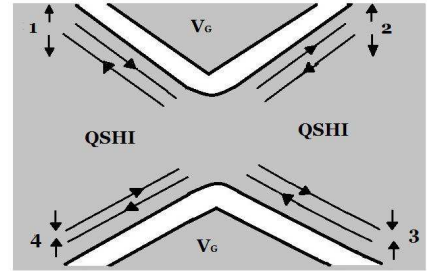}
 \caption{(Color online) %
 A QPC in a QSHI. The value of the gate voltage is greater
 than its critical value so that the point contact is open.}
 \label{hlpc1}
\end{center}
\end{figure}

It is important to notice that in determining the phase diagram of the QPC in a QSHI, the two-particle
scattering processes, which are naively regarded as less relevant than the single-particle one, play an
important role. It is therefore interesting and important to realize a direct experimental probe of
these two-particle scattering processes. One way to achieve this goal is to analyze the current noise of
the tunnel junction. A pioneering work along this direction has been done recently in Ref. \onlinecite{TS}.
There, via the cumulant generating function, the nonequilibrium spin-resolved tunnel current and its
correlations at zero frequency are obtained by the perturbation theory in the tunneling strength.
Particularly, the competition between the single-particle and the two-particle scattering processes has
already been seen in the zero-frequency tunnel current noise. Further, fermionic Hanbury-Brown and Twiss
(HBT) correlations between spin-up and spin-down tunnel currents are also examined in the same work, and it
is shown that only the two-particle tunnelings contribute to the HBT correlations.

On the other hand, it was argued in the study of the noise measurement in the edge states of the fractional
quantum Hall (FQH) liquid that more information can be extracted from the noise correlations of the currents
in the terminals than from the noise in the tunnel current flowing through the junction.\cite{CFW,Sander} In
particular, fractional charge quasi-particle excitations have been suggested\cite{CFW,Sander,safi,SV,nayak}
and measured\cite{Saminadayar,heiblum} in transport through FQH liquids (or chiral LLs) as well as in the
nonchiral LLs.\cite{hur1} Recently, it has been proposed that fractional charge quasi-particle excitations
induced by electron interactions exist and may be probed in the helical edge states of the QSHI.\cite{hur2}
It is therefore of great interest and fundamental importance to investigate further this issue which is
relatively less studied. Furthermore, in addition to the zero-frequency shot noise, it has been suggested
that even more information is stored in the finite-frequency current noise, such as the quantum statistics
of the quasi-particle excitations,\cite{safi,SV,nayak} the dynamics of correlations,\cite{CFW} and the role
of electron interactions.

Motivated by these observations, in the present work, we investigate the current noise of two weakly
coupled helical LLs in a generic four-terminal setup in the presence of a finite bias $V$ between the top
and bottom edges of the point contact when it is open, as shown in Fig. \ref{hlpc1}. In particular, we
extend the earlier studies on the zero-frequency tunnel current noise in Ref. \onlinecite{TS} in two
directions. First, instead of studying the correlations of the tunnel current directly, we investigate
currents in the four terminals and the associated noise spectra. Second, instead of just calculating the
noise spectrum at zero frequency, we also obtain the noise spectra at finite frequency. As we
mentioned above, there is certain important physical information about this system, which is not probed
directly by the tunnel current noise at zero frequency, that can be revealed through this approach. In particular,
we show that the Fano factor obtained through the correlations of the currents in the terminals will
depend on the LL parameter in a way exhibiting the fractional charge carried by the elementary excitations
in the helical LL. Therefore, in addition to reproducing some results of Ref. \onlinecite{TS}, our work may
provide a complementary point of view on the physics of the helical LL in such a four-terminal setup.

Throughout this work, we assume that our system is far away from the quantum critical point so that the
perturbative calculation becomes reliable. Further, when the point contact is pinched off, the corresponding
noise spectra can be obtained by an appropriate duality transformation.\cite{TK} Our main results obtained
via the Keldysh perturbation theory\cite{Kel} are shown in Figs. \ref{hli2}--\ref{fano02}. Naively, the
two-particle scattering processes seem to be more irrelevant than the single-particle one. However, we find that
the current and noise spectrum may be dominated by them, depending on the value of the LL parameter,\cite{LOB}
as shown in Figs. \ref{hli2} - \ref{hls21}. To reveal the competition between the single-particle and the
two-particle scattering processes clearly, we calculate the Fano factors for the auto- and cross correlations
of currents in the terminals. We find that they depend on the LL parameter and the relative strength of the
tunneling processes, which can also be viewed as an indirect probe on the possible fractional charges in a
helical liquid. What we find here is quite different from the corresponding one for tunneling between chiral
edge states in the quantum Hall
liquid.\cite{CFW,KF2} Moreover, the Fano factors for the auto- and cross correlations approach different
values in the zero-bias limit, depending on the LL parameter. As we discuss below, this result follows
from the entanglement of the right and left movers in the final states for different scattering processes.

The rest of the paper is organized as follows. In Sec. \ref{h1}, we set up the model to fix our notation. The
calculations on the currents and noise spectra are summarized in Sec. \ref{h2}. These results and comparison
with the previous work are discussed in Sec. IV. The last section is devoted to conclusions.

\section{Model}
\label{h1}

At low energies, the system in Fig.1 can be described by the
Hamiltonian: $H=H_0+\delta H$,
where
\begin{equation}
 H_0=\! \sum_{i=1}^4 \! \int^{+\infty}_0 \! \! dx{\mathcal H}_0^{(i)} \ ,
 \label{pshlh1}
\end{equation}
with
\begin{eqnarray}
 {\mathcal H}_0^{(i)} \! \! &=& \! \! iv_F(\psi^{\dagger}_{i,\mbox{{\small in}}}
 \partial_x\psi_{i,\mbox{{\small in}}}-\psi^{\dagger}_{i,\mbox{{\small out}}}
 \partial_x\psi_{i,\mbox{{\small out}}})+u_2J_{i,\mbox{{\small in}}}
 J_{i,\mbox{{\small out}}} \nonumber \\
 \! \! & & \! \! +\frac{u_4}{2}(J_{i,\mbox{{\small in}}}J_{i,\mbox{{\small in}}}
 +J_{i,\mbox{{\small out}}}J_{i,\mbox{{\small out}}}) \ , \label{pshlh11}
\end{eqnarray}
and $\delta H$ being defined below.
Here $\psi_{i,\mbox{{\small in}}}$, $\psi_{i,\mbox{{\small out}}}$ are a time
reversed pair of fermion fields with opposite spin, which propagate toward and away
from the junction, $v_F$ is the bare Fermi velocity, and the $u_2$, $u_4$ terms are
forward scattering. As pointed out in Ref. \onlinecite{HKC}, $H_0$ can be mapped
onto the Hamiltonian of spin-$1/2$ fermions. To proceed, we define the spin-$1/2$
fermion fields as
\begin{eqnarray}
 \psi_{R\uparrow}(x) &=& \! \left\{\begin{array}{cc}
 \psi_{2,\mbox{{\small out}}}(x) & x>0 \\
 & \\
 \psi_{1,\mbox{{\small in}}}(-x) & x<0
 \end{array}\right. , \nonumber \\
 \psi_{R\downarrow}(x) &=& \! \left\{\begin{array}{cc}
 \psi_{3,\mbox{{\small out}}}(x) & x>0 \\
 & \\
 \psi_{4,\mbox{{\small in}}}(-x) & x<0
 \end{array}\right. , \nonumber \\
 \psi_{L\uparrow}(x) &=& \! \left\{\begin{array}{cc}
 \psi_{3,\mbox{{\small in}}}(x) & x>0 \\
 & \\
 \psi_{4,\mbox{{\small out}}}(-x) & x<0
 \end{array}\right. , \nonumber \\
 \psi_{L\downarrow}(x) &=& \! \left\{\begin{array}{cc}
 \psi_{2,\mbox{{\small in}}}(x) & x>0 \\
 & \\
 \psi_{1,\mbox{{\small out}}}(-x) & x<0
 \end{array}\right. . \label{pshl2}
\end{eqnarray}
In terms of $\psi_{L\sigma}$ and $\psi_{R\sigma}$, where
$\sigma =\uparrow,\downarrow=+,-$, $H_0$ can be written as
\begin{equation}
 H_0=\! \int^{+\infty}_{-\infty} \! \! dx{\mathcal H}_0 \ , \label{pshlh12}
\end{equation}
where
\begin{eqnarray}
 {\mathcal H}_0 \! \! &=& \! \! \sum_{\sigma}\! \left[iv_0 \! \left(
 \psi^{\dagger}_{L\sigma}\partial_x\psi_{L\sigma}-\psi^{\dagger}_{R\sigma}
 \partial_x\psi_{R\sigma}\right) \! +u_2J_{L\sigma}J_{R-\sigma}\right]
 \nonumber \\
 \! \! & & \! \! +\frac{u_4}{2}\sum_{\sigma}(J_{L\sigma}J_{L\sigma}
 +J_{R\sigma}J_{R\sigma}) \ . \label{pshlh13}
\end{eqnarray}
It is now clear that Eq. (\ref{pshlh12}) is nothing but the Hamiltonian of
the spin-$1/2$ fermions.

Using the bosonization formulas,\cite{bos}
\begin{eqnarray*}
 \psi_{L\sigma} &=& \frac{1}{\sqrt{2\pi a_0}}\eta_{\sigma}
 e^{-i\sqrt{4\pi}\phi_{L\sigma}} \ , \\
 \psi_{R\sigma} &=& \frac{1}{\sqrt{2\pi a_0}}\eta_{\sigma}
 e^{i\sqrt{4\pi}\phi_{R\sigma}} \ ,
\end{eqnarray*}
and defining the bosonic fields
\begin{eqnarray*}
 \Phi_{\sigma}=\phi_{L\sigma}+\phi_{R\sigma} \ , ~~
 \Theta_{\sigma}=\phi_{L\sigma}-\phi_{R\sigma} \ ,
\end{eqnarray*}
where $a_0$ is the short-distance cutoff, $H_0$ becomes
\begin{equation}
 H_0=\! \sum_{\alpha =c,s}\frac{v_{\alpha}}{2} \! \int^{+\infty}_{-\infty} \! \!
 dx: \! \left[K_{\alpha}(\partial_x\Theta_{\alpha})^2 \! +\frac{1}{K_{\alpha}}
 (\partial_x\Phi_{\alpha})^2\right] \! : \ , \label{pshlh14}
\end{equation}
where $K_c=K$, $K_s=1/K$, $v_c=v=v_s$,
\begin{eqnarray*}
 \Phi_c=\frac{1}{\sqrt{2}}(\Phi_++\Phi_-) \ , ~~
 \Phi_s=\frac{1}{\sqrt{2}}(\Phi_+-\Phi_-) \ ,
\end{eqnarray*}
and similar expressions for $\Theta_{c,s}$. The Klein factors $\eta_{\sigma}$ are
usually chosen to satisfy $\eta_+\eta_-=i$. When spin is conserved at the junction,
there are four fixed points.\cite{KFFN} These include the perfectly transmitting
(CC) limit, in which both charge and spin conduct, the perfectly reflecting (II)
limit, in which both charge and spin are insulating, and the mixed fixed points,
denoted by CI (IC), in which charge (spin) is perfectly transmitting and spin
(charge) is perfectly reflecting. According to the analysis in Refs. \onlinecite{HKC}
and \onlinecite{TK}, the CC and II phases are separated by a quantum phase transition
line by varying the gate voltage. This occurs when $1/2<K<2$. This is the region that
we study.

With the help of $\Phi_c$ and $\Phi_s$, $\delta H$ in the CC limit is given by
\begin{eqnarray}
 \delta H \! \! \! &=& \! \! \! \left[v_ee^{i\sqrt{2\pi}\Phi_c(0)}+{\mathrm H.c.}
 \right] \! \cos{\! \left[\sqrt{2\pi}\Phi_s(0)\right]} \nonumber \\
 \! \! \! & & \! \! + \! \left[v_{\rho}e^{i\sqrt{8\pi}\Phi_c(0)}+{\mathrm H.c.}
 \right] \! +v_{\sigma}\cos{\! \left[\sqrt{8\pi}\Phi_s(0)\right]} , ~~~~~~~~
 \label{pshlh2}
\end{eqnarray}
In terms of the fermion fields, the various terms in $\delta H$ can be written as
\begin{eqnarray*}
 v_e: & & \psi^{\dagger}_{L\sigma}\psi_{R\sigma}+{\mathrm H.c.} \ , \\
 v_{\rho}: & & \psi^{\dagger}_{L\uparrow}\psi_{R\uparrow}
 \psi^{\dagger}_{L\downarrow}\psi_{R\downarrow}+{\mathrm H.c.} \ , \\
 v_{\sigma}: & & \psi^{\dagger}_{L\uparrow}\psi_{R\uparrow}
 \psi^{\dagger}_{R\downarrow}\psi_{L\downarrow}+{\mathrm H.c.} \ .
\end{eqnarray*}
Thus, $v_e$ represents the backscattering of a single electron across the point
contact, $v_{\rho}$ denotes the process involving the tunneling of spin (not charge)
between the top and bottom edges, and $v_{\sigma}$ represents the process involving
the tunneling of charge $2e$ between the top and bottom edges. For the weak potential
strength, the three terms are irrelevant when $1/2<K<2$. In general, higher-order
terms could also be included. However, those terms are less relevant. It suffices to
keep the terms in Eq. (\ref{pshlh2}) to determine the phase diagram. In the following,
we compute the noise spectrum in the CC limit to study the effects of the
two-particle scattering processes.

\section{Nonequilibrium current and noise}
\label{h2}

To analyze the transport properties of this system, we apply a voltage bias $V$ between
the upper and lower edges of the point contact. In such a
case, $H_0$ becomes
\begin{eqnarray*}
 H_0 &=& \! \sum_{i=1}^4 \! \int^{+\infty}_0 \! \! dx{\mathcal H}_0^{(i)} \!
 - \! \sum_{i=1,2} \! \int^{+\infty}_0 \! \! dx\mu_+(
 J_{i,\mbox{{\small in}}}+J_{i,\mbox{{\small out}}}) \\
 & & -\! \sum_{i=3,4} \! \int^{+\infty}_0 \! \! dx\mu_-(
 J_{i,\mbox{{\small in}}}+J_{i,\mbox{{\small out}}}) \\
 &=& \! \int^{+\infty}_{-\infty} \! \! dx \! \left[{\mathcal H}_0 \! -\mu_+
 (J_{R\uparrow}+J_{L\downarrow})-\mu_-(J_{R\downarrow}+J_{L\uparrow})\right]
 ,
\end{eqnarray*}
where $\mu_+-\mu_-=-eV$. (Here we assume that the charge carried by an electron
is $-e$.) To proceed, it is convenient to move the dependence on the chemical
potentials to $\delta H$. This is achieved by the time-dependent gauge transformation:
(throughout the calculations, we set $\hbar=1$.)
\begin{eqnarray*}
 \psi_{R\uparrow}(\psi_{L\downarrow}) &\rightarrow& e^{i\mu_+t}
 \psi_{R\uparrow}(\psi_{L\downarrow}) \ , \\
 \psi_{R\downarrow}(\psi_{L\uparrow}) &\rightarrow& e^{i\mu_-t}
 \psi_{R\downarrow}(\psi_{L\uparrow}) \ ,
\end{eqnarray*}
leading to $\delta H=\sum_{i=1}^3 \! \delta H_i$, where
\begin{eqnarray}
 \delta H_1 &=& \! \left[v_ee^{i\sqrt{2\pi K_c}\tilde{\Phi}_c(t,0)}
 +{\mathrm H.c.}\right] \nonumber \\
 & & \times\cos{\! \left[\sqrt{2\pi K_s}\tilde{\Phi}_s(t,0)-\omega_0t\right]}
 , \nonumber \\
 \delta H_2 &=& v_{\rho}e^{i\sqrt{8\pi K_c}\tilde{\Phi}_c(t,0)}+{\mathrm H.c.} \ ,
 \nonumber \\
 \delta H_3 &=& v_{\sigma}\cos{\! \left[\sqrt{8\pi K_s}\tilde{\Phi}_s(t,0)-2\omega_0t\right]}
 . \label{hls11}
\end{eqnarray}
Here $\tilde{\Phi}_{\alpha}=\Phi_{\alpha}/\sqrt{K_{\alpha}}$ and $\omega_0=eV$. The $\omega_0$
dependence of the various terms reflects the numbers of transferred charges involved in the
corresponding process.

Let $\hat{J}_i$ denote the particle current operator flowing into terminal $i$. Then,
we have
\begin{eqnarray*}
 \hat{J}_1(t,x_1) &=& J_{1,\mbox{{\small in}}}(t,-x_1)-J_{1,\mbox{{\small out}}}
 (t,-x_1) \\
 &=& J_{R\uparrow}(t,x_1)-J_{L\downarrow}(t,x_1) \ , \\
 \hat{J}_2(t,x_2) &=& J_{2,\mbox{{\small in}}}(t,x_2)-J_{2,\mbox{{\small out}}}
 (t,x_2) \\
 &=& J_{L\downarrow}(t,x_2)-J_{R\uparrow}(t,x_2) \ , \\
 \hat{J}_3(t,x_3) &=& J_{3,\mbox{{\small in}}}(t,x_3)-J_{3,\mbox{{\small out}}}
 (t,x_3) \\
 &=& J_{L\uparrow}(t,x_3)-J_{R\downarrow}(t,x_3) \ , \\
 \hat{J}_4(t,x_4) &=& J_{4,\mbox{{\small in}}}(t,-x_4)-J_{4,\mbox{{\small out}}}
 (t,-x_4) \\
 &=& J_{R\downarrow}(t,x_4)-J_{L\uparrow}(t,x_4) \ ,
\end{eqnarray*}
where $x_1,x_4<0$ and $x_2,x_3>0$. In terms of the bosonic fields, $\hat{J}_i$ can
be written as
\begin{eqnarray}
 & & \hat{J}_1=\sqrt{\frac{K_s}{2\pi}}\partial_x\tilde{\Phi}_s-\frac{1}
     {\sqrt{2\pi K_c}}\partial_x\tilde{\Theta}_c=-\hat{J}_2 \ , \nonumber \\
 & & \hat{J}_3=\sqrt{\frac{K_s}{2\pi}}\partial_x\tilde{\Phi}_s+\frac{1}
     {\sqrt{2\pi K_c}}\partial_x\tilde{\Theta}_c=-\hat{J}_4 \ . \label{qcpj1}
\end{eqnarray}
where $\tilde{\Theta}_{\alpha}=\sqrt{K_{\alpha}}\Theta_{\alpha}$. The current flowing into
terminal $i$ is given by $I_i=-ev_F\langle\hat{J}_i\rangle$. According to our convention,
$I_i$ is positive when the current flows out of the terminal.

The noise spectrum is defined by
\begin{equation}
 S_{ij}(\omega; ,x, x^\prime)\equiv e^2v_F^2 \! \! \int^{+\infty}_{-\infty} \! \! dte^{i\omega t}\langle
 \{\Delta\hat{J}_i(t, x),\Delta\hat{J}_j(0, x^\prime)\}\rangle \ , \label{noise1}
\end{equation}
where $\Delta\hat{J}_i=\hat{J}_i-\langle\hat{J}_i\rangle$. We would like to calculate
$I_i$ and $S_{ij}$ in terms of the perturbative expansion in the tunneling amplitude
$v_l$ ($l=e,\rho,\sigma$) within the Keldysh formalism.\cite{Kel} We shall see later
that $\langle\hat{J}_i\rangle=O(|v_l|^2)$. Thus, to order of $|v_l|^2$, $S_{ij}$ can
be written as
\begin{eqnarray*}
 S_{ij}(\omega; x, x^\prime)=e^2v_F^2 \! \int^{+\infty}_{-\infty} \! \! dte^{i\omega t}\langle\{
 \hat{J}_i(t, x),\hat{J}_j(0, x^\prime)\}\rangle \ .
\end{eqnarray*}

The perturbative calculations of the current and noise spectrum can be straightly performed
using the Keldysh functional integral formulation, as was done in Ref. \onlinecite{CFW} for
tunneling between the chiral LLs. To the order of $|v_l|^2$, the currents at zero temperature
are given by
\begin{widetext}
\begin{equation}
 I_1(t)=-\frac{e}{2}\mbox{sgn}(\omega_0) \! \left[|v_e|^2\mbox{Re}({\mathcal A})
 |\omega_0|^{K+1/K-1}+|v_{\sigma}|^2\mbox{Re}({\mathcal B}_s)|2\omega_0|^{4/K-1}\right]
 \! =I_2(t)=-I_3(t)=-I_4(t) \ , \label{hli1}
\end{equation}
and the noise spectra at zero temperature are given by
\begin{eqnarray}
 S_{ii}(\omega) &=& e^2 \! \left\{\frac{K}{\pi}|\omega|+ \! \left[\frac{1-K^2}{2}
 |v_e|^2\mbox{Im}({\mathcal A})|\omega_0|^{K+1/K-1}+|v_{\sigma}|^2\mbox{Im}({\mathcal B}_s)
 |2\omega_0|^{4/K-1}\right] \! \sin{\! \left(\frac{2|\omega x|}{v}\right)}\right. \nonumber
 \\
 & & +\frac{|v_e|^2}{8}(1-K^2) \! \left({\mathcal A}e^{i\frac{2|\omega x|}{v}}
 +{\mathrm C.c.}\right) \! \! \left[|\omega+\omega_0|^{K+1/K-1}+|\omega-\omega_0|^{K+1/K-1}
 \right] \nonumber \\
 & & +\frac{|v_{\sigma}|^2}{4} \! \left({\mathcal B}_se^{i\frac{2|\omega x|}{v}}
 +{\mathrm C.c.}\right) \! \! \left[|\omega+2\omega_0|^{4/K-1}+|\omega-2\omega_0|^{4/K-1}
 \right] \nonumber \\
 & & +|v_e|^2\mbox{Re}({\mathcal A})(|\omega_0|-|\omega|)^{K+1/K-1} \! \left[
 \sin^2{\! \left(\frac{\omega x}{v}\right)}+K^2\cos^2{\! \left(\frac{\omega x}{v}\right)}
 \right] \! \theta (|\omega|)\theta (|\omega_0|-|\omega|) \nonumber \\
 & & +2|v_{\sigma}|^2\mbox{Re}({\mathcal B}_s)(|2\omega_0|-|\omega|)^{4/K-1}
 \sin^2{\! \left(\frac{\omega x}{v}\right)}\theta (|\omega|)\theta (|2\omega_0|-|\omega|)
 \nonumber \\
 & & \! \left.-2K^2|v_{\rho}|^2 \! \left({\mathcal B}_ce^{i\frac{2|\omega x|}{v_c}}
 +{\mathrm C.c.}\right) \! |\omega|^{4K-1}\right\} , \label{hls1}
\end{eqnarray}
with $i=1,2,3,4$, and
\begin{eqnarray}
 S_{12}(\omega) &=& -e^2 \! \left\{\frac{K|\omega|}{\pi}
 \cos{\! \left(\frac{2\omega x}{v}\right)}-\! \left[\frac{1+K^2}{2}|v_e|^2\mbox{Im}
 ({\mathcal A})|\omega_0|^{K+1/K-1}+|v_{\sigma}|^2\mbox{Im}({\mathcal B}_s)|2\omega_0|^{4/K-1}
 \right] \! \sin{\! \left(\frac{2|\omega x|}{v}\right)}\right. \nonumber \\
 & & -\frac{|v_e|^2}{8}(1+K^2) \! \left({\mathcal A}e^{i\frac{2|\omega x|}{v}}
 +{\mathrm C.c.}\right) \! \! \left[|\omega+\omega_0|^{K+1/K-1}+|\omega-\omega_0|^{K+1/K-1}
 \right] \nonumber \\
 & & -\frac{|v_{\sigma}|^2}{4} \! \left({\mathcal B}_se^{i\frac{2|\omega x|}{v}}
 +{\mathrm C.c.}\right) \! \! \left[|\omega+2\omega_0|^{4/K-1}+|\omega-2\omega_0|^{4/K-1}
 \right] \nonumber \\
 & & +|v_e|^2\mbox{Re}({\mathcal A})(|\omega_0|-|\omega|)^{K+1/K-1} \! \left[\frac{i}{2}
 \sin{\! \left(\frac{2\omega |x|}{v}\right)} \! +K^2\cos^2{\! \left(\frac{\omega x}{v}\right)}
 \right] \! \theta (|\omega|)\theta (|\omega_0|-|\omega|) \nonumber \\
 & & +i|v_{\sigma}|^2\mbox{Re}({\mathcal B}_s)(|2\omega_0|-|\omega|)^{4/K-1}
 \sin{\! \left(\frac{2\omega |x|}{v}\right)}\theta (|\omega|)\theta (|2\omega_0|-|\omega|)
 \nonumber \\
 & & \! \left.-2K^2|v_{\rho}|^2 \! \left({\mathcal B}_ce^{i\frac{2|\omega x|}{v}}
 +{\mathrm C.c.}\right) \! |\omega|^{4K-1}\right\} , \label{hls2}
\end{eqnarray}
where $S_{ii}(\omega)=S_{ii}(\omega;x,x)$, $S_{ij}(\omega)=S_{ij}(\omega;x, -x)$ for $i\neq j$,
\begin{eqnarray*}
 \mbox{Re}({\mathcal A})=\frac{\pi a_0^{K+1/K}}{v^{K+1/K}\Gamma (K+1/K)} \ , & &
 \mbox{Im}({\mathcal A})=-\frac{\pi a_0^{K+1/K}\tan{[\pi (K+1/K)/2]}}
 {v^{K+1/K}\Gamma (K+1/K)} \ , \\
 \mbox{Re}({\mathcal B}_s)=\frac{\pi a_0^{4/K}}{v^{4/K}\Gamma (4/K)} \ , & &
 \mbox{Im}({\mathcal B}_s)=-\frac{\pi a_0^{4/K}\tan{(2\pi/K)}}{v^{4/K}\Gamma (4/K)}
 \ ,
\end{eqnarray*}
and
\begin{eqnarray*}
 \! \left({\mathcal A}e^{i\frac{2|\omega x|}{v}}+{\mathrm C.c.}\right) \! &=&
 \frac{2\pi a_0^{K+1/K}}{v^{K+1/K}\Gamma (K+1/K)} \! \left\{
 \cos{\! \left(\frac{2|\omega x|}{v}\right)} \! +\tan{\! \left[\frac{\pi}{2}(K+1/K)\right]}
 \sin{\! \left(\frac{2|\omega x|}{v}\right)}\right\} , \\
 \! \left({\mathcal B}_{\alpha}e^{i\frac{2|\omega x|}{v}}+{\mathrm C.c.}\right) \! &=&
 \frac{2\pi a_0^{4K_{\alpha}}}{v^{4K_{\alpha}}\Gamma (4K_{\alpha})} \! \left[
 \cos{\! \left(\frac{2|\omega x|}{v}\right)} \! +\tan{(2\pi K_{\alpha})}
 \sin{\! \left(\frac{2|\omega x|}{v}\right)}\right] .
\end{eqnarray*}
\end{widetext}
On account of current conservation, the tunneling current $I_t$ is given by $I_t=-(I_1+I_2)=I_3+I_4$.
This has been verified by directly calculating $I_t=-e\langle\hat{J}_t\rangle$ through the tunnel
current operator
\begin{eqnarray}
 \hat{J}_t &=& -\! \left[v_ee^{i\sqrt{2\pi}\Phi_c(0)}+{\mathrm H.c.}\right] \!
 \sin{\! \left[\sqrt{2\pi}\Phi_s(0)\right]} \nonumber \\
 & & -2v_{\sigma}\sin{\! \left[\sqrt{8\pi}\Phi_s(0)\right]} . \label{jt1}
\end{eqnarray}

\section{Results and Discussions}
\label{h3}

\begin{figure}
\begin{center}
 \includegraphics[width=0.9\columnwidth]{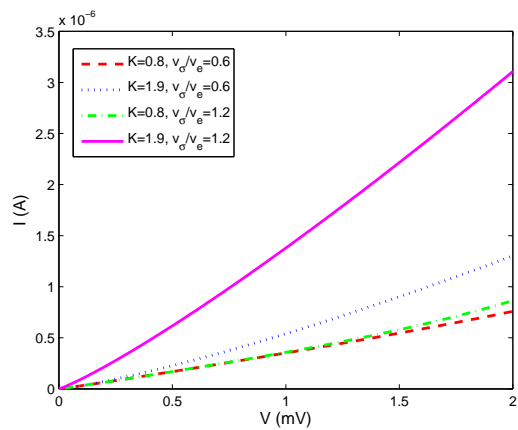}
 \caption{(Color online) %
 Dependence of the current $I=|I_i|$ on the bias $V$. We use the parameters
 $a_0=10^{-7}$m, $v=5.5\times 10^5$ m/s, and $|v_e|=\hbar v/a_0$.}
 \label{hli2}
\end{center}
\end{figure}

We now discuss our results. The result for the current is shown in Fig. \ref{hli2}. Although what
we are considering is the nonequilibrium transport, it is interesting to see that the
dependence of each term in Eq. (\ref{hli1}) on the bias follows from the scaling dimension
of each scattering process. [The scaling dimensions of the $v_e$, $v_{\rho}$, and
$v_{\sigma}$ terms about the LL fixed point are $\Delta_e=(K+1/K)/2$, $\Delta_{\rho}=2K$,
and $\Delta_{\sigma}=2/K$, respectively.] We notice that the $v_{\rho}$ term completely
disappears in Eq. (\ref{hli1}) because it does not involve net charge transport. Albeit that
this term plays an important role in determining the phase diagram, our perturbative
calculations show that its effects on the electrical transport can only be probed through
the current correlations at finite frequency, as shown in Eqs. (\ref{hls1}) and (\ref{hls2}).

\begin{figure}
\begin{center}
 \includegraphics[width=0.9\columnwidth]{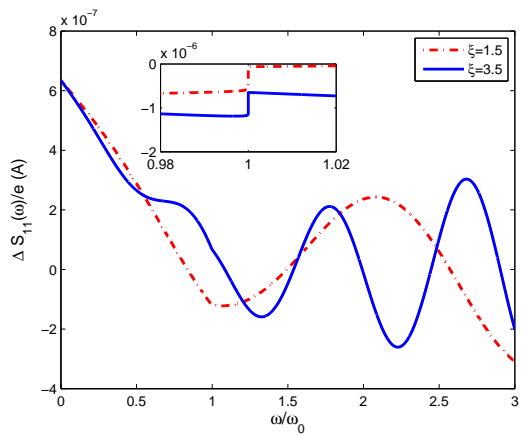}
 \caption{(Color online) %
 Dependence of $\Delta S_{11}=S_{11}-S_{11}^{(0)}$ on the frequency $\omega$ at $K=0.85$, where
 $S^{(0)}_{11}$ is the noise spectrum in the absence of tunneling, that is, $v_l=0$. As explained in
 the text, the singularity at $\omega=\omega_0$ is much stronger than that at $\omega=2\omega_0$
 in the region where the single-particle scattering is dominated ($1/2<K<\sqrt{3}$), and thus a
 clear structure can be seen in the figure near $\omega=\omega_0$. To reveal the singularity at
 $\omega=\omega_0$, a plot of $d\Delta S_{11}/d\omega$ around $\omega=\omega_0$ is shown in the
 inset. We use the parameters $a_0=10^{-7}$m, $v=5.5\times 10^5$ m/s,
 $|v_e|=\hbar v/a_0=|v_{\sigma}|=|v_{\rho}|$, and $\xi=x/(\hbar v/\omega_0)$.}
 \label{s11w1}
\end{center}
\end{figure}

\begin{figure}
\begin{center}
 \includegraphics[width=0.9\columnwidth]{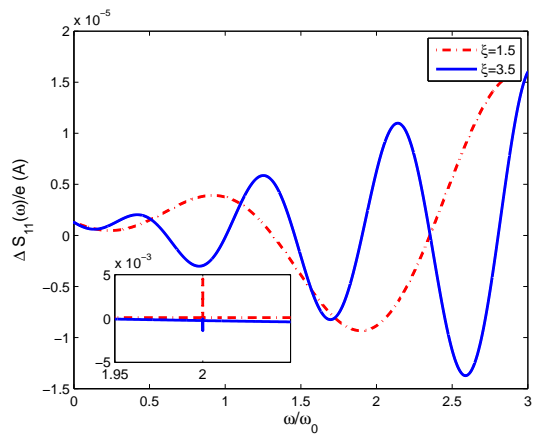}
 \caption{(Color online) %
 Dependence of $\Delta S_{11}=S_{11}-S_{11}^{(0)}$ on the frequency $\omega$ at $K=1.85$, where
 $S^{(0)}_{11}$ is the noise spectrum in the absence of tunneling, that is, $v_l=0$. As explained in
 the text, the singularity at $\omega=2\omega_0$ is relatively weaker than that at $\omega=\omega_0$
 in the region where the two-particle scattering is dominated ($\sqrt{3}<K<2$). To reveal the former,
 a plot of $d^2\Delta S_{11}/d\omega^2$ around $\omega=2\omega_0$ is shown in the inset. We use the
 parameters $a_0=10^{-7}$ m, $v=5.5\times 10^5$ m/s, $|v_e|=\hbar v/a_0=|v_{\sigma}|=|v_{\rho}|$, and
 $\xi=x/(\hbar v/\omega_0)$.}
 \label{s11w2}
\end{center}
\end{figure}

After examining the voltage dependence of the current, we now turn to the voltage and frequency
dependence of the noise spectrum. Since the behavior of $\mbox{Re}\{S_{12}(\omega)\}$ is similar
to that of $S_{11}(\omega)$, we just plot the frequency dependence of $\Delta S_{11}=S_{11}-S_{11}^{(0)}$
in Figs. \ref{s11w1} and \ref{s11w2}, where $S^{(0)}_{11}$ is the noise spectrum in the absence
of tunneling, that is, $v_l=0$. We would like to emphasize a few features. First of all, the noise
spectrum at finite frequency is sensitive to the position of the probe, with the overall
oscillatory behavior determined by the sine or cosine functions. Next, in addition to the
singularity at $\omega=0$, $S_{ij}(\omega)$ also exhibits singularities at $\omega =\omega_0$ and
$\omega=2\omega_0$, corresponding to the single-particle and two-particle tunneling, respectively.
We expect that these singularities remain intact even by taking into account the non-perturbative
effects.\cite{CFW} For $1/2<K<\sqrt{3}$ (the regime dominated by the single-particle scattering $v_e$
term), the singularity at $\omega=\omega_0$ reveals itself in the guise of a cusp in
$dS_{11}/d\omega$, as shown in the inset of Fig. \ref{s11w1}. However, the sub-leading singularity
at $\omega=2\omega_0$ in this region can only be seen in the higher-order derivatives of $S_{11}$
due to its higher powers. For example, near $2\omega_0$, $S_{11}\sim |\omega-2\omega_0|^{3.71}$ at
$K=0.85$; hence, one can see the singular behavior at $\omega=2\omega_0$ at least in the fourth-order
derivative $d^4S_{11}/d\omega^4$. On the other hand, for $\sqrt{3}<K<2$ (the regime dominated by
the two-particle scattering $v_\sigma$ term), the singular behaviors of $S_{11}$ around
$\omega=\omega_0$ and $\omega=2\omega_0$ are not clear as shown in Fig. \ref{s11w2} due to its high
powers in $|\omega-\omega_0|$ and $|\omega-2\omega_0|$ in this region. Nevertheless, the dominated
singular behavior at $\omega=2\omega_0$ can still be revealed through the second-order derivative
$d^2S_{11}/d\omega^2$, as shown in the inset of Fig. \ref{s11w2}. These qualitative features shown
in $S_{ij}(\omega)$ can be used in future experiments to probe the dynamical current correlations
of the helical LLs in the presence of both single-particle and two-particle scattering.

\begin{figure}
\begin{center}
 \includegraphics[width=0.9\columnwidth]{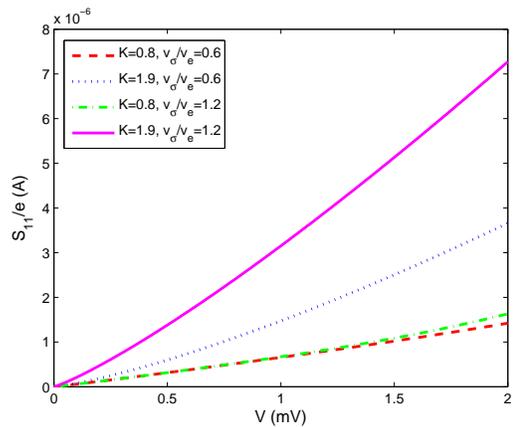}
 \caption{(Color online) %
 Dependence of the autocorrelation at zero frequency $S_{11}(0)$ on the bias $V$.
 We use the parameters $a_0=10^{-7}$m, $v=5.5\times 10^5$ m/s, and $|v_e|=\hbar v/a_0$.}
 \label{hls12}
\end{center}
\end{figure}

\begin{figure}
\begin{center}
 \includegraphics[width=0.9\columnwidth]{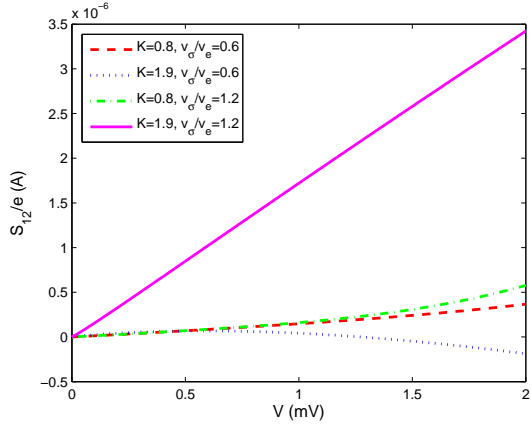}
 \caption{(Color online) %
 Dependence of the cross correlation at zero frequency $S_{12}(0)$ on the bias
 $V$. We use the parameters $a_0=10^{-7}$m, $v=5.5\times 10^5$ m/s, and $|v_e|=\hbar v/a_0$.}
 \label{hls21}
\end{center}
\end{figure}

In contrast to the tunneling between chiral LLs in a similar four-terminal configuration where
only the cross correlations for the chiral currents depend on the position of the probe,\cite{CFW}
both the auto- and the cross correlations in the present case are sensitive to the position of the
probe $x$. It turns out that their zero-frequency limits are the most robust measurements of
fluctuations in the present situation because the resulting expressions in this limit are
independent of the position of the probe. The dependence of $S_{11}(0)$ and $S_{12}(0)$ on the bias
$V$ is shown in Figs. \ref{hls12} and \ref{hls21}. We notice that the bias dependence of each term
in $S_{11}(0)$ and $S_{12}(0)$ also follows from the scaling dimension of each scattering process.
Therefore, among the three scattering terms, only one scattering process will dominate the behavior
of $S_{ij}$ at low bias, depending on the LL parameter $K$, though the introduction of two-particle
scattering will, in general, enhance the strength of the current correlation. It follows from Eqs.
(\ref{hls1}) and (\ref{hls2}) that $S_{ij}(0)$ at low bias are dominated by the single-particle
scattering term (the $v_e$ term) at $1/2<K<\sqrt{3}$, while at $\sqrt{3}<K<2$ it is the $v_{\sigma}$
term that is dominant. A direct consequence of this is that, as shown in Figs. \ref{hls12} and
\ref{hls21}, the increase of the magnitude of $S_{ij}(0)$ with $K=1.9$ at low bias is much larger
than the one of $S_{ij}(0)$ with $K=0.8$ for the same amount of the increment of the ratio
$|v_{\sigma}/v_e|$. A similar situation also occurs for the current, as shown in Fig. \ref{hli2}.

It should be noted here that, at $\sqrt{3}<K<2$, the exponents $K+1/K-1$ and $4/K-1$ are
numerically quite close to one another. This indicates that both the $v_e$ and the $v_{\sigma}$
terms will contribute significantly to $S_{ij}(0)$, except for the case when the bias is
extremely low. For example, at $K=1.9$, we need to take $V$ to be about $0.01$ meV in order to
see that the contribution of the $v_{\sigma}$ term is $10$ times larger than that of the $v_e$
term, assuming that $|v_{\sigma}/v_e|=O(1)$. Therefore, a better way to reveal the competition
between the single-particle and the two-particle scattering processes is to investigate the
Fano factor, which is defined by
\begin{equation}
 F_{ij}(V)=\frac{S_{ij}(0)}{2e|I|} \ . \label{fano1}
\end{equation}
Since the Fano factor is directly related to the charge fluctuations in the terminals, as we
shall demonstrate later, it contains the information about the fractional charge excitations
in the helical LL.
From the noises and currents calculated above, we have
\begin{eqnarray}
 F_{ii}(V) &=& \frac{1}{2} \! \left[\frac{1+K^2+2\eta |v_{\sigma}/v_e|^2
 |\omega_0|^{3/K-K}}{1+\eta |v_{\sigma}/v_e|^2|\omega_0|^{3/K-K}}\right] ,
 \nonumber \\
 F_{12}(V) &=& \frac{1}{2} \! \left[\frac{1-K^2+2\eta |v_{\sigma}/v_e|^2
 |\omega_0|^{3/K-K}}{1+\eta |v_{\sigma}/v_e|^2|\omega_0|^{3/K-K}}\right] ,
 \label{fano2}
\end{eqnarray}
where
\begin{eqnarray*}
 \eta &\equiv& 2^{4/K-1}\frac{\mbox{Re}({\mathcal B}_s)}{\mbox{Re}({\mathcal A})}
 \\
 &=& 2^{4/K-1} \! \left(\frac{a_0}{v}\right)^{\! 3/K-K} \! \frac{\Gamma (K+1/K)}
 {\Gamma (4/K)} 
\end{eqnarray*}
is a nonuniversal constant.

\begin{figure}
\begin{center}
 \includegraphics[width=0.9\columnwidth]{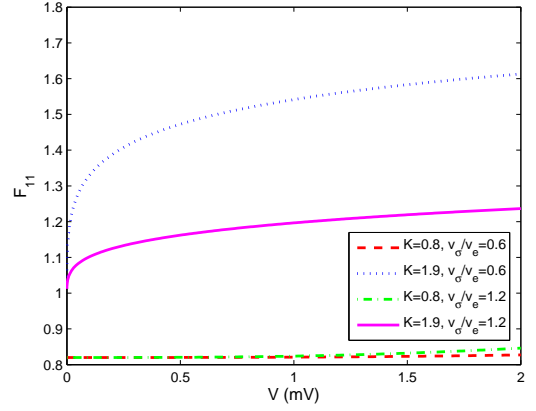}
 \caption{(Color online) %
 Dependence of $F_{11}$ on the bias $V$. We use the parameters $a_0=10^{-7}$ m
 and $v=5.5\times 10^5$ m/s.}
 \label{fano01}
\end{center}
\end{figure}

\begin{figure}
\begin{center}
 \includegraphics[width=0.9\columnwidth]{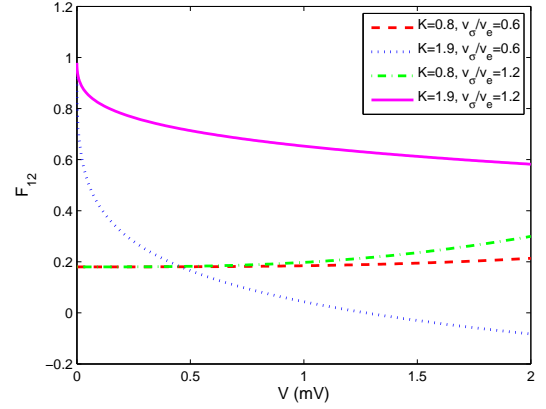}
 \caption{(Color online) %
 Dependence of $F_{12}$ on the bias $V$. We use the parameters $a_0=10^{-7}$ m
 and $v=5.5\times 10^5$ m/s.}
 \label{fano02}
\end{center}
\end{figure}

Since the $v_{\rho}$ dependence completely disappears in the zero-frequency limit, $F_{ij}$
is very sensitive to the single ratio $|v_{\sigma}/v_e|$. In general, $F_{ij}(V)$ consists of
terms exhibiting a power law in $V$ with exponents related to the scaling dimension of each
scattering process. We plot $F_{11}$ and $F_{12}$ as functions of the bias $V$ in Figs.
\ref{fano01} and \ref{fano02}. The effects of the $v_e$ and the $v_{\sigma}$ terms are
disentangled at the zero-bias limit. By taking $V\rightarrow 0$, we find that
\begin{equation}
 F_{ii}(0)=\! \left\{\begin{array}{cc}
 \frac{1+K^2}{2} & 1/2<K<\sqrt{3} \, , \\
 & \\
 1 & \sqrt{3}<K<2\, , 
 \end{array}\right.  \label{fano3}
\end{equation}
and
\begin{equation}
 F_{12}(0)=\! \left\{\begin{array}{cc}
 \frac{1-K^2}{2} & 1/2<K<\sqrt{3}\, ,  \\
 & \\
 1 & \sqrt{3}<K<2\, .
 \end{array}\right.  \label{fano32}
\end{equation}
Note that in the region where the $v_{\sigma}$ term dominates, $F_{ij}(0)$ takes the classical
Schottky result, while it depends on the LL parameter $K$ in the region where the single-particle
tunneling is dominant. As a by-product, the behaviors of $F_{ij}(V)$ may provide us with a way to
measure the value of $K$ for the helical liquid.

In Ref.~\onlinecite{PGL} (see also Refs.~\onlinecite{hur1,hur2,CGDM}), it was shown that when a charge
is injected into a LL, it will break up into two counterpropagating---left-moving and 
right-moving---quasiparticles carrying fractional charges.
We now apply this idea to interpret our
results [Eqs. (\ref{fano3}) and (\ref{fano32})]. Since the effects of the $v_e$ and the $v_{\sigma}$
terms are disentangled in the limit $V\rightarrow 0$, we consider this limit first. Without loss of
generality, we assume that $V>0$. Then the $v_e$ term implies a single-electron tunneling from the
bottom edge to the top one, whereas the $v_{\sigma}$ term implies the simultaneous tunneling of a
spin-up electron and a spin-down electron from the bottom edge to the top one. The former ($v_e$)
process generates the following state:
\begin{eqnarray*}
 & & \! \sum_{\sigma}\Psi^{\dagger}_{\sigma}(x=0)|O_{LL}\rangle \\
 & & = \! \sum_{\sigma}\psi^{\dagger}_{R\sigma}(x=0)|O_{LL}\rangle+ \! \sum_{\sigma}
     \psi^{\dagger}_{L\sigma}(x=0)|O_{LL}\rangle \ ,
\end{eqnarray*}
while the state produced by the latter ($v_\sigma$) is
\begin{eqnarray*}
 \psi^{\dagger}_{R\uparrow}(x=0)\psi^{\dagger}_{L\downarrow}(x=0)|O_{LL}\rangle \ ,
\end{eqnarray*}
where $|O_{LL}\rangle$ denotes the ground state of the LL. In the above, the terms with higher scaling
dimensions are neglected. To proceed, we define the new chiral bosonic fields
\begin{eqnarray*}
 \phi_{\alpha l}=\frac{1}{2}(\tilde{\Phi}_{\alpha}+\tilde{\Theta}_{\alpha}) \ , ~
 \phi_{\alpha r}=\frac{1}{2}(\tilde{\Phi}_{\alpha}-\tilde{\Theta}_{\alpha}) \ ,
\end{eqnarray*}
where $\alpha =c,s$. $\phi_{\alpha l}$ and $\phi_{\alpha r}$ describe the elementary excitations of
the spin-$1/2$ LL propagating with speed $v$ along the left and the right directions, respectively.
In terms of $\phi_{\alpha l}$ and $\phi_{\alpha r}$, we may define the chiral fields carrying a unit
of U($1$) charge\cite{PGL, Safi}:
\begin{equation}
 \tilde{\psi}_{cl}=\exp{\! \left[-i\sqrt{\frac{2\pi}{K}}\phi_{cl}\right]} \ , ~~
 \tilde{\psi}_{cr}=\exp{\! \left[i\sqrt{\frac{2\pi}{K}}\phi_{cr}\right]} \ . \label{u13}
\end{equation}
Then we have for the single-particle ($v_e$) process
\begin{eqnarray}
 & & \! \sum_{\sigma}\Psi^{\dagger}_{\sigma}(x=0)|O_{LL}\rangle \label{u11} \\
 & & =\! \left[\tilde{\psi}_{cl}^{\dagger}(x=0)\right]^{Q_-} \! \left[\tilde{\psi}_{cr}^{\dagger}
     (x=0)\right]^{Q_+} \! O_{s1}(x=0)|O_{LL}\rangle \nonumber \\
 & & +\! \left[\tilde{\psi}_{cl}^{\dagger}(x=0)\right]^{Q_+} \! \left[\tilde{\psi}_{cr}^{\dagger}
     (x=0)\right]^{Q_-} \! O_{s2}(x=0)|O_{LL}\rangle \ , \nonumber
\end{eqnarray}
and for the two-particle ($v_\sigma$) process we obtain
\begin{eqnarray}
 & & \psi^{\dagger}_{R\uparrow}(x=0)\psi^{\dagger}_{L\downarrow}(x=0)|O_{LL}\rangle \label{u12}
     \\
 & & =\tilde{\psi}_{cl}^{\dagger}(x=0)\tilde{\psi}_{cr}^{\dagger}(x=0)O_{s3}(x=0)|O_{LL}\rangle
     \ , \nonumber
\end{eqnarray}
where
\begin{eqnarray*}
 Q_{\pm}=\frac{1\pm K}{2} \ ,
\end{eqnarray*}
and
\begin{eqnarray*}
 O_{s1} &=& e^{i\frac{\pi}{16K}(1-K^2)} \! \sum_{\sigma}\frac{\eta_{\sigma}}{\sqrt{2\pi a_0}}
 e^{-i\sqrt{\frac{\pi}{2}}\sigma(\Phi_s-\Theta_s)} \ , \\
 O_{s2} &=& e^{i\frac{\pi}{16K}(1-K^2)} \! \sum_{\sigma}\frac{\eta_{\sigma}}{\sqrt{2\pi a_0}}
 e^{i\sqrt{\frac{\pi}{2}}\sigma(\Phi_s+\Theta_s)} \ , \\
 O_{s3} &=& \frac{\eta_{\uparrow}\eta_{\downarrow}}{2\pi a_0}e^{i\frac{(1+K)\pi}{4K}}
 e^{-i\sqrt{\frac{\pi}{2}}(\Phi_s-\Theta_s)}e^{-i\sqrt{\frac{\pi}{2}}(\Phi_s+\Theta_s)} \ .
\end{eqnarray*}
Since the operators $O_{s1}$, $O_{s2}$, and $O_{s3}$ are charge neutral, by focusing only on
the charge states we may reexpress Eqs. (\ref{u11}) and (\ref{u12}) as
\begin{eqnarray*}
 \sum_{\sigma}\Psi^{\dagger}_{\sigma}(x=0)|O_{LL}\rangle &\sim& |Q_+,Q_-\rangle +|Q_-,Q_+\rangle
 \ , \\
 \psi^{\dagger}_{R\uparrow}(x=0)\psi^{\dagger}_{L\downarrow}(x=0)|O_{LL}\rangle &\sim& |1,1\rangle
 \ ,
\end{eqnarray*}
where $|Q_l,Q_r\rangle$ denotes the charge state in which the left and the right movers carry
charge $Q_l$ and $Q_r$, respectively. Both the above expressions can be understood as the consequence of
fractionalization of charge upon its injection into a LL as discussed in Ref. \onlinecite{PGL}.

In one spatial dimension, the current fluctuations amount
to the measurement of charge fluctuations. Accordingly, we get
\begin{eqnarray*}
 & & S_{ii}(0)\propto Q_+^2+Q_-^2=\frac{1+K^2}{2} \ , \\
 & & S_{12}(0)\propto 2Q_+Q_-=\frac{1-K^2}{2} \ ,
\end{eqnarray*}
when the $v_e$ term dominates, while for the $v_{\sigma}$ term being dominant, $S_{ij}(0)$ is proportional to
a $K$-independent constant.
From the above analysis, we see that the dependence of $F_{ij}(0)$ on the LL parameter $K$ follows
from the fact that the final state of the single-particle scattering is an entangled state of the
left- and the right-mover carrying fractional charge $Q_{\pm}e$. On the other hand, the classical
Schottky result arises from the final state of the two-particle scattering, which is a direct
product state of the left and the right mover both carrying charge $-e$. (This state is not a
direct product state of the single-electron states because here the left- and the right-mover
carry fractional spins, $\pm 1/K$ in units of $\hbar/2$.) At finite bias, both the $v_e$ and the
$v_{\sigma}$ terms will contribute to the current and the current noise so that the Fano factor
depends on the ratio $|v_{\sigma}/v_e|$.

We may now compare our results with the main results in Ref. \onlinecite{TS}. First of all, in
terms of current conservation, that is, $I_t=-I_1(0^-)-I_2(0^+)$, we can obtain the tunnel current
noise at finite frequency:
\begin{eqnarray}
 S_t(\omega) &\equiv& e^2 \! \int^{+\infty}_{-\infty} \! \! dt~e^{i\omega t}\langle\{\Delta\hat{J}_t(t),
 \Delta\hat{J}_t(0)\}\rangle \nonumber \\
 &=& S_{11}(\omega;0^-,0^-)+S_{22}(\omega;0^+,0^+) \nonumber \\
 & & +S_{12}(\omega;0^-,0^+)+S_{21}(\omega;0^+,0^-) \ , \label{jt2}
\end{eqnarray}
where $\Delta\hat{J}_t=\hat{J}_t-\langle\hat{J}_t\rangle$ and $\hat{J}_t$ is given by Eq. (\ref{jt1}).
Inserting Eqs. (\ref{hls1}) and (\ref{hls2}) into Eq. (\ref{jt2}), we find that
\begin{eqnarray}
 & & S_t(\omega) \label{jt22} \\
 & & =e^2 \! \left[|v_e|^2\mbox{Re}({\mathcal A}) \! \left(|\omega+\omega_0|^{K+\frac{1}{K}-1}
     +|\omega-\omega_0|^{K+\frac{1}{K}-1}\right)\right. \nonumber \\
 & & \left.+2|v_{\sigma}|^2\mbox{Re}({\mathcal B}_s) \! \left(|\omega+2\omega_0|^{4/K-1}
     +|\omega-2\omega_0|^{4/K-1}\right)\right] , \nonumber
\end{eqnarray}
to $O(|v_l|^2)$. Equation (\ref{jt22}) can also be obtained by a direction calculation using $\hat{J}_t$
defined in Eq. (\ref{jt1}). The zero-frequency limit, $S_t(0)$, coincides with the result in Ref.
\onlinecite{TS} [see Eq. (21) of Ref. \onlinecite{TS}]. It should be emphasized that although we may obtain
the total tunnel current noise in Ref. \onlinecite{TS} from our $S_{ij}$, the reverse is not true. This
is simply because the currents in the four terminals, $I_1$, $I_2$, $I_3$, and $I_4$, cannot be expressed
by the tunnel currents $I_{t\uparrow}$ and $I_{t\downarrow}$, where $I_{t\uparrow}$ and $I_{t\downarrow}$
are spin-up and spin-down tunnel currents, respectively. The other way to see the difference between the
two approaches can be seen from the Fano factor for the tunnel current, which is defined as
$F_t(V)=S_t(0)/(2e|I_t|)$. To $O(|v_l|^2)$, we have
\begin{equation}
 F_t(V)=\frac{1+2\eta |v_{\sigma}/v_e|^2|\omega_0|^{3/K-K}}{1+\eta |v_{\sigma}/v_e|^2|\omega_0|^{3/K-K}}
 , \label{jt23}
\end{equation}
leading to
\begin{eqnarray*}
 F_t(0)=\! \left\{\begin{array}{cc}
 1 & 1/2<K<\sqrt{3}\, ,  \\
 & \\
 2 & \sqrt{3}<K<2\, , 
 \end{array}\right. 
\end{eqnarray*}
in the zero-bias limit. We see that the Fano factor in the zero-bias limit $F_t(0)$, corresponding
to the effective quasiparticle charge transporting in the tunneling process, exhibits the
classical Schottky result: For the single-particle-process-dominated region ($1/2<K<\sqrt{3}$),
$F_t(0) = 1$ (corresponding to charge $e$), whereas for the region dominated by the two-particle process ($\sqrt{3}<K<2$),
$F_t(0)=2$ (corresponding to charge 2$e$). This must be the case since only electrons can tunnel between the two edges. By contrast,
the currents in the terminals consist of quasiparticles which may carry fractional charge; the Fano
factors for the currents in the terminals can therefore be used to detect the fractionally charged
elementary excitations in the helical LL, as discussed. Hence, our work contains unique information
about the nature of the fractional charge elementary excitations of the helical edge states, which
is not seen in the tunnel current noises as studied in Ref. \onlinecite{TS}. On the other hand, as
shown in Ref. \onlinecite{TS}, the cross correlation between the spin-up and spin-down tunnel
currents can be used to study the fermionic HBT correlations. Since the currents $I_i$ studied here
are not spin-resolved, our present results can not be used to address such an issue, and it is
beyond the scope of our present work.

It is interesting to notice that in the case of tunneling between the chiral LLs, the Fano factor
takes the classical Schottky result.\cite{CFW} In the present case, however, the Fano factor is a
function of the LL parameter $K$ even in the absence of the two-particle tunneling. Similar results
also occur for tunneling into a nanotube.\cite{CGDM} Hence, our work offers a way to distinguish
the spin-$1/2$ LL from the chiral LL.

Finally, we would like to point out that, as noticed in Ref. \onlinecite{TK}, there exists a
duality relation between the CC and the II limits. Therefore, the noise spectrum in the II limit
in the presence of a bias between the left and right edges can be obtained from our results by
interchanging the LL parameters of the charge and spin modes, that is, $K\leftrightarrow 1/K$.

\section{Conclusions}

To summarize, we have studied the current and the noise spectrum of a four-terminal QPC in the
QSHI at finite bias. Special emphasis is put on the fractional charge quasiparticle excitations
shown in the noise correlations of the currents in the terminals (in contrast to the tunneling
current noise spectrum) and examining how the single-particle and the two-particle scattering
processes compete with each other. Via the Keldysh perturbative approach, we obtained noise
spectra of the currents in the terminals, which are, in general, sensitive to the ratios of the
tunneling strength and consist of terms exhibiting power law in bias voltage $V$ with the
exponents determined from the scaling dimension of each scattering process. We find that both
auto- and cross correlations of the noise spectra $S_{ij}(\omega)$ are sensitive to the
positions of the probe with an overall oscillatory behavior. Meanwhile, $S_{ij}(\omega)$ exhibits
singularities at $\omega=\omega_0$ and $\omega=2\omega_0$, corresponding to the single-particle
and the two-particle scattering processes, respectively. It is a unique feature of the helical LL
that the two-particle scattering process dominates the electrical transport at low bias in some
parameter regime. The observation of the corresponding singularity in the finite-frequency noise
spectra is a direct probe of this mechanism.

In addition to revealing the main characteristics of the current correlations at finite frequency,
we also point out the difference between the noise spectra of the helical LL and the chiral LL.
The correlations between the currents in the terminals studied in this paper furnish us with
important information about the fractionally charged elementary excitations in the helical LL.
In particular, we find from the Fano factors of the currents in the terminals that the fractional
charge excitations show up in the single-particle scattering dominated regime ($1/2<K<\sqrt{3}$),
whereas the classical Schottky result is obtained in the two-particle scattering dominated regime
($\sqrt{3}<K<2$). We provide further analytical understanding of these results via the idea of
charge fractionalization in LLs with an electron injection as shown in Refs. \onlinecite{CGDM,hur1,hur2}.
Note that this information cannot be extracted from the previous study on the tunnel current
noise.\cite{TS} In fact, we have calculated the Fano factor $F_t(V=0)$, corresponding to the
effective charge transporting through the junction, and found that the result in the zero-bias
limit is nothing but the classical Schottky result for both the single-particle- and the 
two-particle-tunneling-process-dominated regimes. This is in sharp contrast to our results for the Fano factors
of the currents in the terminals. Therefore, our results offer a useful guide for the experimental
identification of the helical LL, and thus the interaction effects in the QSHI.

Recently we became aware of the work by Souquet and Simon,~\cite{PSimon} which has partial overlap with the present work. The finite-frequency tunneling current noise was calculated, and
both singularities associated with the one-particle ($\omega=eV$) and the two-particle ($\omega=2eV$) processes were also found in their results.
\acknowledgments

The works of Y.-W. Lee and Y.L. Lee are supported by the National Science
Council of Taiwan under Grants No. NSC 99-2112-M-029-006-MY3 and
No. NSC 98-2112-M-018-003-MY3, respectively.
C.-H Chung acknowledges support by the NSC Grant No. 98-2112-M-009-010-MY3,
No.101-2628-M-009-001-MY3, the NCTU-CTS, NCTS, and the MOE-ATU program of Taiwan.


\end{document}